\documentclass[aps,prb,amsmath,amssymb,
 reprint]{revtex4-1}
\usepackage{graphicx}
\usepackage{dcolumn}
\usepackage{bm}
\usepackage{color}
\usepackage{graphicx}
\usepackage[english]{babel}
\usepackage{color}
\usepackage{booktabs,longtable}
\usepackage{natbib}
\usepackage{xspace}
\usepackage{textgreek} 
\usepackage{upgreek} 
\usepackage{mathrsfs}
\usepackage[colorlinks]{hyperref}
\hypersetup{
                pdfstartview={FitH},
                linkcolor=blue,
                citecolor=blue,
                filecolor=blue,
                urlcolor=blue
}

\newcommand{\fig}[2]{\textcolor{blue}{Fig.~\ref{Fig#1} #2)}}
\newcommand{\figsimple}[1]{\textcolor{blue}{Fig.~\ref{Fig#1}}}

\newcommand{\degreeC}{$^{\circ}$C\xspace}
\newcommand{\degree}{$^{\circ}$\xspace}
\newcommand{\kelv}{$~\textup{K}$\xspace}
\newcommand{\volt}{$~\textup{V}$\xspace}
\newcommand{\tesla}{$~\textup{T}$\xspace}
\newcommand{\uA}{$~\text{\textmu A}$\xspace}
\newcommand{\um}{$~\text{\textmu m}$\xspace}

\begin{document}

\begin{sloppypar}

\title{Large Rashba unidirectional magnetoresistance in the Fe/Ge(111) interface states}

\author{T. Guillet}
\affiliation{Univ. Grenoble Alpes, CEA, CNRS, Grenoble INP, IRIG-SPINTEC, 38000 Grenoble, France}
\author{C. Zucchetti}
\affiliation{LNESS-Dipartimento di Fisica, Politecnico di Milano, Piazza Leonardo da Vinci 32, 20133 Milano, Italy}
\author{A. Marty}
\affiliation{Univ. Grenoble Alpes, CEA, CNRS, Grenoble INP, IRIG-SPINTEC, 38000 Grenoble, France}
\author{G. Isella}
\affiliation{LNESS-Dipartimento di Fisica, Politecnico di Milano, Piazza Leonardo da Vinci 32, 20133 Milano, Italy}
\author{C. Vergnaud}
\affiliation{Univ. Grenoble Alpes, CEA, CNRS, Grenoble INP, IRIG-SPINTEC, 38000 Grenoble, France}
\author{Q. Barbedienne}
\affiliation{Unit\'e Mixte de Physique, CNRS, Thales, Univ. Paris-Sud, Universit\'e Paris-Saclay, 91767, Palaiseau, France}
\author{H. Jaffr\`es}
\affiliation{Unit\'e Mixte de Physique, CNRS, Thales, Univ. Paris-Sud, Universit\'e Paris-Saclay, 91767, Palaiseau, France}
\author{N. Reyren}
\affiliation{Unit\'e Mixte de Physique, CNRS, Thales, Univ. Paris-Sud, Universit\'e Paris-Saclay, 91767, Palaiseau, France}
\author{J.-M. George}
\affiliation{Unit\'e Mixte de Physique, CNRS, Thales, Univ. Paris-Sud, Universit\'e Paris-Saclay, 91767, Palaiseau, France}
\author{A. Fert}
\affiliation{Unit\'e Mixte de Physique, CNRS, Thales, Univ. Paris-Sud, Universit\'e Paris-Saclay, 91767, Palaiseau, France}
\author{M. Jamet}
\affiliation{Univ. Grenoble Alpes, CEA, CNRS, Grenoble INP, IRIG-SPINTEC, 38000 Grenoble, France}

\date{\today}

\begin{abstract}
The structure inversion asymmetry at surfaces and interfaces give rise to the Rashba spin-orbit interaction (SOI), that breaks the spin degeneracy of surface or interface states. Hence, when an electric current runs through a surface or interface, this Rashba effect generates an effective magnetic field acting on the electron spin. This provides an additional tool to manipulate the spin state in materials such as Si and Ge that, in their bulk form, possess inversion  symmetry (or lack structural inersion asymmetry). The existence of Rashba states could be demonstrated by photoemission spectroscopy at the interface between different metals and Ge(111) and by spin-charge conversion experiments at the Fe/Ge(111) interface even though made of two light elements. In this work, we identify the fingerprint of the Rashba states at the Fe/Ge(111) interface by magnetotransport measurements in the form of a large unidirectional magnetoresistance of up to 0.1 \%. From its temperature dependence, we find that the Rashba energy splitting is larger than in pure Ge(111) subsurface states. 
\end{abstract}

\maketitle

\section{Introduction}

One of the main promises of spintronics is the implementation of a semiconducting platform where both logic and memory operations could be performed simultaneously, increasing significantly the speed and energy efficiency of such devices. \cite{bohr_cmos_2017,wolf_spintronics_2001} Intensive work has been carried out to effectively generate and detect spin accumulations in semiconductors\cite{van_t_erve_electrical_2007,koo_electrical_2007,suzuki_room-temperature_2011} and metals\cite{jedema_electrical_2001,jedema_electrical_2002} for the last 20 years. 
These operations are usually performed using magnetic tunnel junctions in a so-called lateral spin valve geometry\cite{johnson_interfacial_1985,johnson_spin-injection_1988}. This architecture is very close to the prototypical spin transistor originally proposed by Datta and Das in 1990\cite{datta_electronic_1990}.
In such structure, the manipulation of spins relies on the Larmor precession by applying an external magnetic field \cite{fabian_semiconductor_nodate}, which is incompatible with the very large scale integration paradigm. Therefore, the need for an all-electrical manipulation of the spin orientation is still an on-going field and is currently stimulating the scientific community.

Recently, we reported the observation of a large unidirectional magnetoresistance (UMR) in the group-IV semiconductor Ge(111)\cite{guillet_observation_2020}. The detected magnetoresistance exhibits two characteristic features: it is unidirectional and linear with the applied magnetic field and electrical current\cite{he_bilinear_2018,Zhang2018,he_observation_2018,Rikken1,Rikken2,Bibes2020}. We showed that this UMR originates from the interplay between an external magnetic field and the Rashba pseudo-magnetic field induced by a current passing through the spin-splitted subsurface states of Ge(111). It is understood as a consequence of the strong spin-orbit interaction (SOI) and is a signature of the Rashba effect acting on the spins. We also showed that the effect was orders of magnitude stronger than in high SOI systems like Bi$_2$Se$_3$\cite{he_bilinear_2018} or SrTiO$_3$\cite{he_observation_2018} and could be enchanced or suppressed by applying a gate voltage within a 10\volt range.

Although the intrinsic Ge Rashba spin-splitting of the subsurface states is not strong enough to target room temperature applications, it can be greatly enhanced at the interface between Ge(111) and metals\cite{ohtsubo_spin-polarized_2010,ohtsubo_two-dimensional_2013,aruga_different_2015,yaji_experimental_2015}. Hence, beyond providing a way to manipulate the electron spin state, these Rashba states could also be used to generate and detect spin currents in germanium at room temperature, which represents an interesting concept in the field of semiconductor spintronics.
In this respect, we previously explored the influence of putting a metal, potentially heavy, in contact with germanium in order to induce and enhance the Rashba SOI\cite{oyarzun_evidence_2016,zucchetti_tuning_2018}. In particular, Fe/Ge(111) and Bi/Ge(111) bilayers were extensively studied by a variety of experimental and theoretical techniques. In both systems, we observed spin-charge interconversion at the interface between the two materials due to the presence of Rashba states. 

In this work, we investigate the effect of Rashba states at the Fe/Ge interface on magnetotransport properties. Since Fe and Ge are both conducting materials, the electrical conduction will occur in three parallel channels, exhibiting specific magnetoresistance effects. The Fe thickness was varied from 0 nm to 3 nm by depositing a wedge of Fe by molecular beam epitaxy (MBE) in order study the effect of the Fe/Ge interface.
The magnitude of the UMR sharply decreases when increasing the Fe thickness as a consequence of the current shunting in the ferromagnetic film. Nonetheless, we managed to observe simultaneously the magnetotransport signatures of the ferromagnetic film and the UMR related to the presence of the Rashba states, which represents a promising observation for applications as one could take profit of the strong spin-polarization of the Rashba states to manipulate the magnetic state of the FM film. Interestingly, when increasing the temperature, the UMR decreases slower than in the case of pure Ge. We conclude that the Rashba energy splitting at the Fe/Ge(111) interface is larger than the one in the subsurface states of Ge(111). It indicates that the Rashba-SOI is reinforced by the addition of Fe atoms at the Ge(111) surface. 

\section{Sample preparation}

In this study, we use as a substrate a 2\um -thick Ge/Si(111) film, deposited by low-energy plasma-enhanced chemical vapor deposition (LEPECVD)\cite{gatti_gesige_2014}. The deposition rate was $\approx$4 nm s$^{-1}$ and the substrate temperature was fixed at 500\degreeC. Post-growth annealing cycles have been used to improve the crystal quality. The Ge layer is non intentionally doped with a residual hole carrier concentration $p$ $\approx 2 \times10^{16} \textup{cm}^{-3}$ as measured by Hall effect at room temperature.

The low $p$-doped Ge/Si(111) substrate is subsequently cleaned in acetone and isopropanol in an ultrasonic bath for 5 minutes to remove organic species. Then the substrate is dipped during 30 secondes in a 50 \% hydrofluoric acid solution to remove the native oxide layer and is loaded into the MBE chamber. 
The substrate is then annealed at 850\degreeC during 30 minutes in ultrahigh vacuum (base pressure $\sim10^{-10}$ mbar) to evaporate the remaining Ge oxide and degas remaining adsorbents. The clean Ge surface is then bombarded for 20 minutes using soft argon milling, the argon pressure in the chamber is maintained at 5.50$\times$10$^{-5}$ mbar, the acceleration voltage is 450 V, creating a sample current of 4\uA. This procedure inducing surface disorder, it is followed by a second annealing at 850\degreeC in order to recrystallize the surface.

\begin{figure}[b!]
\begin{center}
\includegraphics[width=0.48\textwidth]{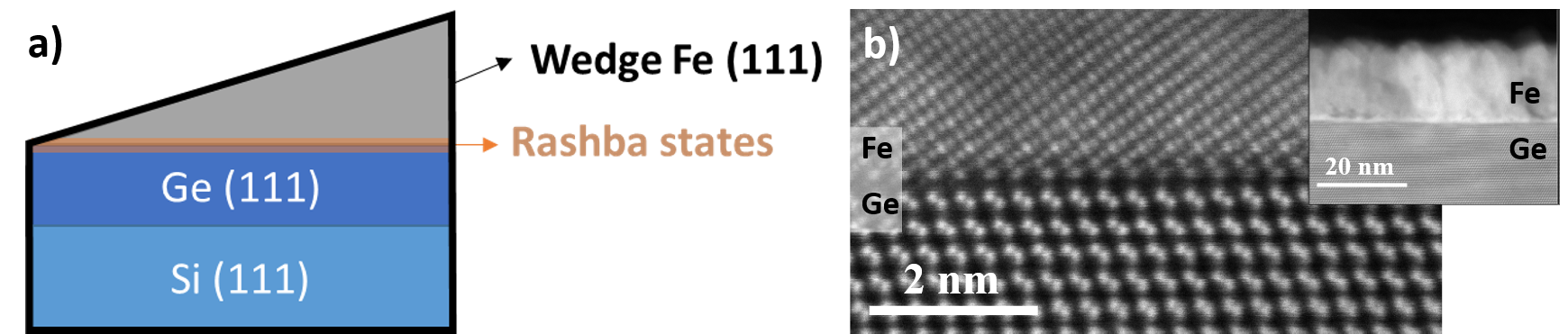} 
\caption{a) Illustration of the Fe/Ge/Si(111) wedge sample. b) High-resolution scanning transmission electron microscopy in cross-section, extracted from Ref.~[\citenum{oyarzun_evidence_2016}].}
\label{Fig1}
\end{center}
\end{figure}

A thickness gradient is obtained by depositing a wedge of Fe at room temperature. The Fe thickness continuously varies from 0 nm to 3 nm as illustrated in \fig{1}{a}. The small lattice mismatch ($\approx$ 1.3 \%) between Ge and bcc-Fe allows the growth of a single crystalline film along the(111) direction with the epitaxial relationship: Fe(111)[11$\bar{2}$] $\vert\vert$ Ge(111)[2$\bar{1}\bar{1}$] \cite{oyarzun_evidence_2016}. The high-resolution scanning transmission electron microscopy image in \fig{1}{b} confirms the epitaxial growth with a very sharp Fe/Ge interface.

The sample is then patterned into 130$\times$35 \textmu m$^2$ Hall bars using standard cleanroom techniques. In short, we use laser lithography and ion beam etching to define the Fe channel and Au(120 nm)/Ti(5 nm) Ohmic contacts are deposited by e-beam evaporation. Ion-coupled plasma is then employed to etch the 2-\um -thick Ge layer in order to limit the current shunting in Ge.

\section{Current distribution in Fe/Ge system}

\fig{2}{a} represents the three parallel conduction channels coexisting in the system. Indeed, the longitudinal zero field resistance $R_{xx,0}$ can be interpreted in the frame of an equivalent parallel resistance, as:

\begin{equation}
R_{xx,0} = R_{\textup{Ge}}\parallel R_{\textup{Fe}}\parallel R_{\textup{2D}}
\end{equation} 

$ R_{\textup{Ge}}$ and $ R_{\textup{Fe}}$ the resistances of bulk Ge and Fe, respectively, and  $R_{\textup{2D}}$ the resistance of the Rashba gas located at the Fe/Ge interface.
The temperature dependence of the resistance of three devices with 1.1 nm, 1.7 nm and 2.3 nm of Fe are reported in \fig{2}{b}. They were measured between 10\kelv and room temperature using a 10\uA AC current at $f = 13.3$ Hz and a lock-in detection. We obtain similar behaviors for the 1.7 nm and 2.3 nm-thick Fe films: a weak temperature dependence exhibiting a minimum value. This behavior was also observed in a Fe/MgO(100) reference sample (not shown). Interestingly, the 1.1 nm-thick Fe film resistance shows a different temperature dependence. The sharp increase at low temperature seems to indicate that the conduction mostly takes place in the thermally activated Ge channel but the resistance increases linearly (and not exponentially) when decreasing the temperature, suggesting that the transport might occur in a third channel: the interface Rashba gas. 

\begin{figure}[t!]
\begin{center}
\includegraphics[width=0.5\textwidth]{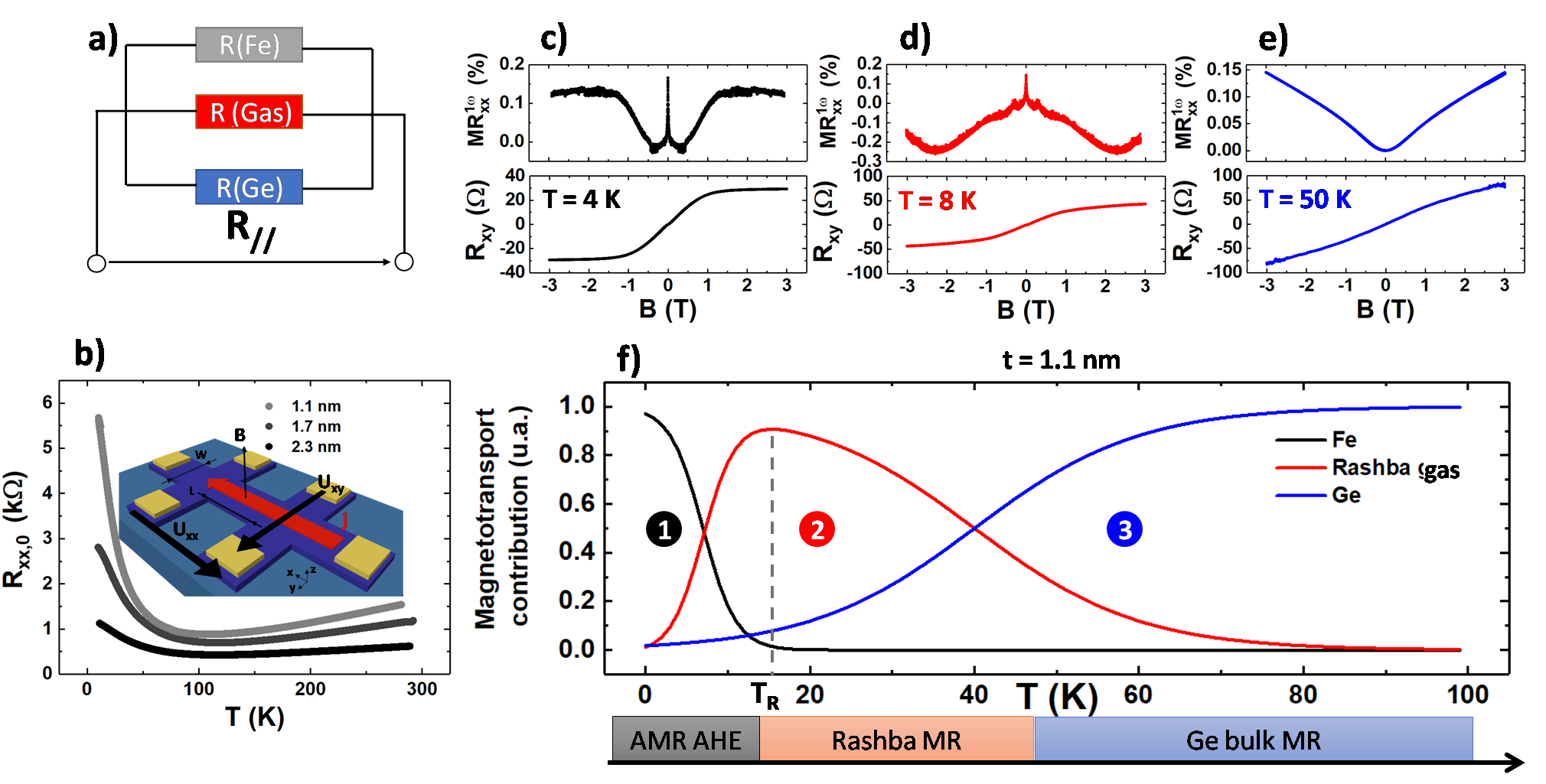} 
\caption{a) Equivalent electrical resistance of the three-channels in the Fe/Ge(111) system. b) Temperature dependence of the resistance for different Fe thicknesses. c-e)  Temperature dependence of the longitudinal magnetoresistance $R_{xx}^{1\omega}$, and transverse resistance $R_{xy}$ using a 10\uA AC current at $f$ = 13.3 Hz and a lock-in detection. f) Qualitative representation of the magnetotransport contributions between the three conduction channels as a function of temperature.}
\label{Fig2}
\end{center}
\end{figure}

To support this interpretation, we performed magnetoresistance (MR) and anomalous Hall effect (AHE) measurements as a function of the sample temperature. \fig{2}{c-e} shows the longitudinal MR curves measured at different temperatures for the 1.1 nm -thick Fe film. The MR is defined as: $\textup{MR}=\frac{R_{xx}(B)-R_{xx}(0)}{R_{xx}(0)}$. At low temperature (4\kelv), the MR is very weak and shows a saturation behavior, this is the manifestation of the anisotropic magnetoresistance (AMR) and indicates that the transport mostly takes place in the Fe film. In the intermediate temperature range, we observe a W-shape MR curve which is no longer related to the saturation field of the Fe film and corresponds to the fingerprint of the Rashba MR\cite{Choi_2015}. At high temperature ($T > 30$\kelv), the characteristic magnetoresistance of Ge is retrieved as the Ge conductivity increases by thermal activation of carriers.

The measurements of the transverse resistance $R_{xy}$ [bottom panel in \fig{2}{c-e}] show a similar behavior: at low temperature, the AHE of Fe is dominant compared to the linear Ge ordinary Hall effect. At higher temperature, the $p$-type conduction in Ge becomes dominant, as indicated by the positive slope for magnetic fields higher than the saturation field of Fe ($\approx~2$\tesla).

Those observations are summarized in \fig{2}{f}, where the contributions to the magnetotransport are qualitatively represented. Note that this figure is only a schematic illustration of the weight of each conduction channel (Fe, Rashba states and Ge) to the longitudinal and transverse magnetoresistances. At low temperature, bulk Ge is insulating and due to the small gap between the top of the Ge valence band and the Rashba states\cite{aruga_different_2015}, magnetotransport mainly takes place into Fe. In an intermediate temperature regime, magnetotransport in the Rashba states is activated and we note $T_R$ the temperature at which it is maximum. Finally, in the high temperature regime, magnetotransport in bulk Ge is dominating. 
All these preliminary observations using first harmonic measurements support our assumption that in addition to the semiconducting (Ge) and metallic (Fe) conduction channels, a third channel is present in the system corresponding to the Rashba gas. 

\section{Angular dependent measurements}

In the Fe conduction channel where the SOI is weak, we observe the AMR reported in \fig{2}{c}, top panel. This MR contribution is current-independent, and quadratic with respect to the applied magnetic field, \textit{i.e.}, by reversing the field direction, the resistance remains unchanged. However, in the Rashba gas, the SOI results in a current-induced effective field that affects the magnetotransport properties. The resulting magnetoresistance is determined by the relative orientation between the applied field and the current-induced effective Rashba field giving rise to a unidirectional magnetoresistance (UMR) contribution. 
The complete theoretical description was developed in Ref.~[\citenum{guillet_observation_2020}]. The corresponding symmetries can be easily found by noting that the total magnetic field acting on the carriers is the vector sum of the external field \textbf{$B$} and the current-induced Rashba field \textbf{$B_E$}. The amplitude of the unidirectional magnetoresistance is expected to be proportional to both the applied magnetic field and the applied current, as the Rashba field is proportional to the current.

\begin{figure}[h!]
\begin{center}
\includegraphics[width=0.5\textwidth]{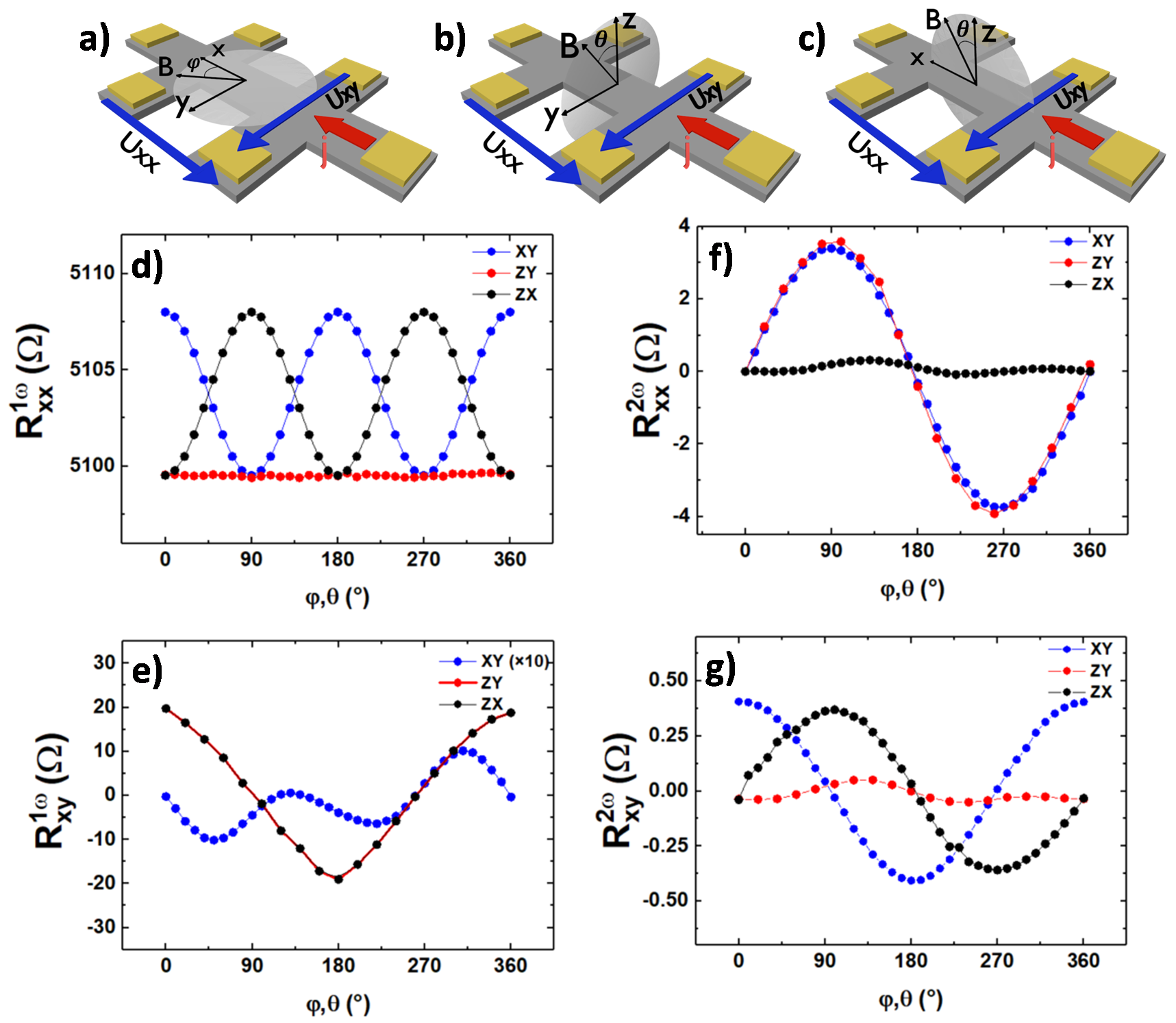} 
\caption{Angular dependences of the longitudinal and transverse signals in the a) ($xy$) b) ($zy$) c) ($zx$) planes. Corresponding angular dependences of d) $R_{xx}^{1\omega}$, e) $R_{xy}^{1\omega}$, f) $R_{xx}^{2\omega}$ and g) $R_{xy}^{2\omega}$, measured at 12\kelv , with an applied magnetic field of 0.5\tesla and a current of 100\uA .}
\label{Fig3}
\end{center}
\end{figure}

In this case, we obtain a resistance term which is linear in current, implying a quadratic dependence of the voltage with the applied current. Since the UMR signals are expected to be rather small due to the current shunting in the Fe layer, we use AC measurements to separate the different harmonics in order to distinguish current-dependent from current-independent resistance contributions.
We apply an AC current $I_{\omega}=I_0\sin(\omega t)$ and simultaneously record the first and second harmonic longitudinal and transverse resistances: $R_{xx}^{1\omega}$, $R_{xx}^{2\omega}$, $R_{xy}^{1\omega}$, $R_{xy}^{2\omega}$. In addition to this harmonic analysis, we use the magnetic field angular dependences to identify all the magnetoresistance contributions.
\fig{3}{a-c} illustrate the different measurement geometries. The applied current is along the $x$ direction and the external magnetic field is applied along (${\theta,\varphi}$) directions, $\theta$  and $\varphi$ being the polar and azimuth angles. The measurement were carried out at 12\kelv using a 100\uA AC current of frequency $f$ = 13.3 Hz, while rotating the sample in the ($xy$), ($zy$) and ($zx$) planes in a uniform external field of 0.5\tesla .
\fig{3}{d} shows the first harmonic longitudinal resistance $R_{xx}^{1\omega}$ angular dependences measured for the 1.1 nm-thick Fe film. We identify the AMR angular dependence: 

\begin{equation}
R_{xx}^{1\omega}=R_{xx}^{1\omega}\vert_{B_x}+\left(R_{xx}^{1\omega}\vert_{B_x}-(R_{xx}^{1\omega}\vert_{B_y}\right)\cos^2\theta\sin^2\varphi
\end{equation}  

Where $R_{xx}^{1\omega}\vert_{B_x}$ (resp. $R_{xx}^{1\omega}\vert_{B_y}$) is the longitudinal resistance for the field oriented along the $x$ (resp. $y$) axis. In the same way, \fig{3}{e} shows the first harmonic transverse resistance $R_{xy}^{1\omega}$ angular dependences measured for the 1.1 nm-thick Fe film. Out-of-plane ($zy$) and ($zx$) scans well correspond to the anomalous Hall effect while the in-plane ($xy$) angular dependence indicates the presence of the planar Hall effect, the transverse counterpart of the AMR. The external field of 0.5\tesla is not strong enough to saturate the Fe magnetization, this results in the sawtooth-like out-of-plane angular dependence in \fig{3}{e}. Overall, this contribution can be expressed as: 

\begin{equation}
R_{xy}^{1\omega}=R_{AHE}\cos\theta+R_{PHE}\sin^2\theta\sin2\varphi
\end{equation}  

\fig{3}{f} shows the second harmonic measurements corresponding to the contributions to the resistance that are current-dependent\cite{avci_unidirectional_2015}. $R_{xx}^{2\omega}$ shows a sine angular dependence with respect to the external field: $R_{xx}^{2\omega}$ changes sign when the external magnetic field is reversed. Regarding the symmetries with respect to the field and current, we call this term unidirectional magnetoresistance (UMR). 
A similar behavior is observed in the second harmonic transverse measurements $R_{xy}^{2\omega}$ as shown in \fig{3}{g}. We define the sine amplitudes as $R_{xx,\Delta}^{2\omega}$ and $R_{xy,\Delta}^{2\omega}$ so that these contributions can be expressed as: 

\begin{equation}
R_{xx}^{2\omega}=R_{xx,\Delta}^{2\omega}\sin\theta\sin\varphi
\end{equation} 

and,

\begin{equation}
R_{xy}^{2\omega}=R_{xy,\Delta}^{2\omega}\sin\theta\cos\varphi
\end{equation} 

Similarly to the case of pure Ge(111), $R_{xy,\Delta}^{2\omega}$ shows the signature of the Nernst effect, we can remove this parasitic contribution from $R_{xx,\Delta}^{2\omega}$ by using the following expression\cite{guillet_observation_2020}: 

\begin{equation} 
R_{\textup{UMR}}^{\Delta}=R_{xx,\Delta}^{2\omega}-Z\times R_{xy,\Delta}^{2\omega}
\end{equation}

\begin{figure}[h!]
\begin{center}
\includegraphics[width=0.5\textwidth]{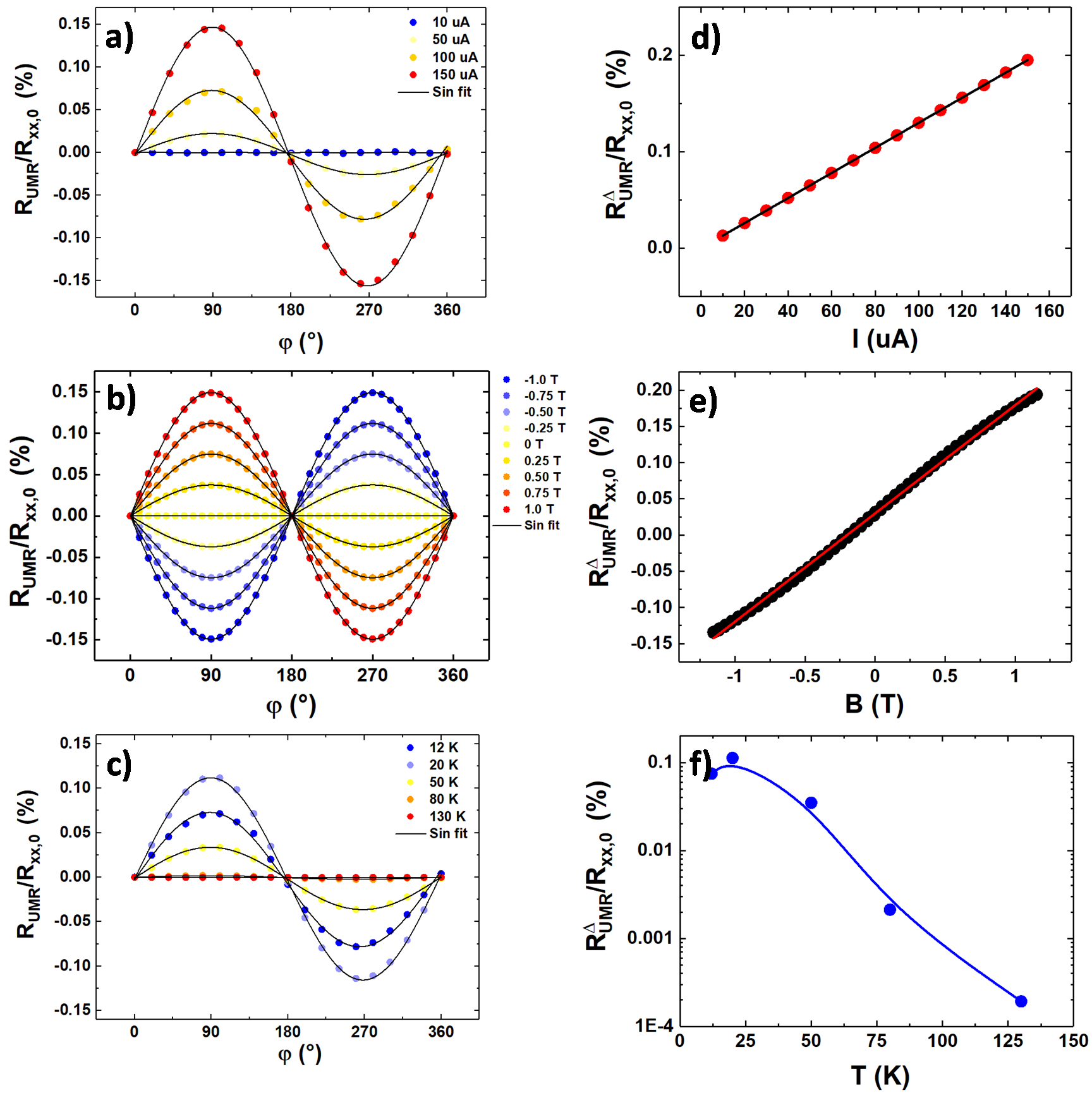} 
\caption{Angular dependence of the second to first harmonic ratio in the ($xy$) plane for  1.1 nm of Fe applying a) different currents (0.5\tesla , 12\kelv), b) different magnetic fields (100\uA , 12\kelv) and c) at different temperatures (0.5\tesla , 100\uA).d-f) Corresponding current, field and temperature dependences for $\varphi$ = 90\degree .}
\label{Fig4}
\end{center}
\end{figure}

Where $Z$ is the aspect ratio of the Hall bar. In \figsimple{4}, we report the dependences of $R_{\textup{UMR}}^{\Delta}$ on the applied current [\fig{4}{a,d}], external magnetic field [\fig{4}{b,e}] and temperature [\fig{4}{c,f}]. The signal is normalized by the zero field longitudinal resistance $R_{xx,0}$ at the corresponding current. The UMR is proportional to the current and magnetic field and thus, follows the symmetries of the current-induced Rashba field. $R_{\textup{UMR}}^{\Delta}/R_{xx,0}$ is maximum at low temperature and sharply decreases with increasing temperature for two reasons: the finite value of the Rashba SOC and the progressive current shunting into the Ge substrate due to carrier thermal activation. Interestingly, the temperature decrease is slightly slower than in pure Ge(111)\cite{guillet_observation_2020}. Hence the Rashba energy splitting in the electronic states at the Fe/Ge(111) interface is larger than in the Ge(111) subsurface states. It indicates that the SOC has been reinforced by the addition of Fe atoms at the Ge(111) surface.

At low temperature, the system is equivalent to a bilayer composed of the Rashba gas and the metallic ferromagnet. Therefore, one would expect to detect the unidirectional spin Hall magnetoresistance effect (USMR)\cite{avci_unidirectional_2015}: $R_{xx,\textup{USMR}}^{2\omega}\propto\textbf{I}\times\textbf{M}$). However this effect is supposed to be two to three orders of magnitude smaller. UMR and USMR effects share the same current and angular dependences, but USMR should follow the Fe magnetization, resulting in the observation of an hysteresis loop at low magnetic field. \fig{4}{e} shows that this contribution cannot be resolved in our system using these measurements conditions.

\begin{figure}[h!]
\begin{center}
\includegraphics[width=0.5\textwidth]{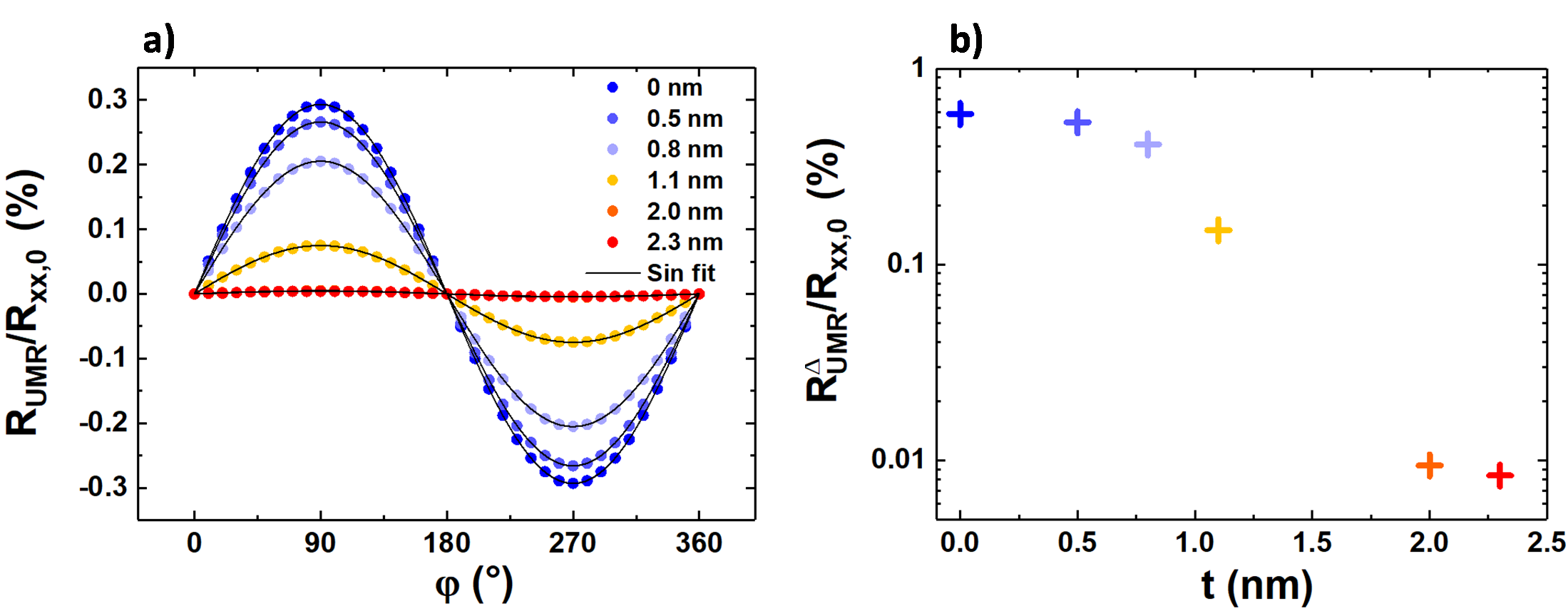} 
\caption{a) Angular dependence of the second to first harmonic ratio in the ($xy$) plane for the different Fe thicknesses measured at 12\kelv , 0.5\tesla and 100\uA . b) Corresponding profile for $\varphi$ = 90\degree .}
\label{Fig5}
\end{center}
\end{figure}

Finally, we present the thickness dependence of $R_{\textup{UMR}}^{\Delta}/R_{xx,0}$ [\figsimple{5}]. The measurements were carried out at 12\kelv , applying an external magnetic field of $B = 0.5$\tesla and an AC excitation current of 100\uA . The UMR intensity decreases as the Fe film becomes thicker, this can be understood since the fraction of the current shunted in the Fe layer becomes larger. We use a figure of merit $\eta$ to make a comparison with previous results on different systems following defined as: ${\eta=R_{\textup{UMR}}^{\Delta}/(R_{xx,0}\,j\,B)}$. At ${15~\text{K}}$, in the 1.1 nm-thick Fe Hall bar, we obtain ${\eta=1.1\times10^{-7}~\textup{cm}^{2}/(\textup{A\,T})}$ assuming that all the electrical current flows within the spatial extension of the subsurface states (10 Ge atomic layers from Ref.~\onlinecite{aruga_different_2015}). It represents a lower bound since part of the current flows in the Fe film. $\eta$ is 4 times lower than in pure $p$-doped Ge(111) but remains orders of magnitude larger than in other systems previously studied\cite{he_bilinear_2018,he_observation_2018}.

\section{Conclusion}

In conclusion, we could measure simultaneously the magnetoresistance contribution from the ferromagnetic Fe layer and the UMR from the Rashba gas at the Fe/Ge(111) interface. This UMR is orders of magnitude larger than in previously studied systems. Although its amplitude is lower than for pure Ge(111), its temperature decay is slower indicating a stronger Rashba spin-orbit coupling. This coupling can be further enhanced by selecting the most suitable metal to interface with germanium to observe the UMR at room temperature.  

\section{Ackowledgements}
The authors acknowledge the financial support from the ANR project ANR-16-CE24-0017 TOP RISE.

\end{sloppypar}

\end{document}